\documentclass[a4paper]{article}
\usepackage{INTERSPEECH2018}
\usepackage{amsmath,graphicx}
\usepackage{diagbox}
\usepackage{multirow}
\usepackage{color}
\usepackage{graphicx}
\usepackage{cases}
\usepackage{subfig}
\usepackage{booktabs}
\usepackage{boldline}
\usepackage{amsfonts}

\title{Single-channel Speech Dereverberation via Generative Adversarial Training}
\name{Chenxing Li$^{1,2}$, Tieqiang Wang$^{1,2}$, Shuang Xu$^1$, Bo Xu$^1$}
\address{
  $^1$Institute of Automation, Chinese Academy of Sciences, Beijing, P.R.China  \\
  $^2$University of Chinese Academy of Sciences, Beijing, P.R.China}
\email{\{lichenxing2015, wangtieqiang2015, shuang.xu, xubo\}@ia.ac.cn}

\begin{document}
\maketitle
\begin{abstract}
In this paper, we propose a single-channel speech dereverberation system (DeReGAT) based on convolutional, bidirectional long short-term memory and deep feed-forward neural network (CBLDNN) with generative adversarial training (GAT). In order to obtain better speech quality instead of only minimizing a mean square error (MSE), GAT is employed to make the dereverberated speech indistinguishable form the clean samples. Besides, our system can deal with wide range reverberation and be well adapted to variant environments. The experimental results show that the proposed model outperforms weighted prediction error (WPE) and deep neural network-based systems. In addition, DeReGAT is extended to an online speech dereverberation scenario, which reports comparable performance with the offline case.
\end{abstract}

\section{Introduction}
\label{s1}
In an enclosed space, speech signals recorded by receivers will be the mixes of original signals and their delayed and attenuated version, this phenomenon is caused by the reflections from different directions. These various reflections damage speech intelligibility, degrade the performance of source localization or speech recognition and are generally hard to be described by deterministic functions or models. To draw these issues, effective speech dereverberation system should be established.

Many approaches have been proposed for decades: (1). Inverse filtering of the room impulse response (RIR) \cite{neely1979invertibility,naylor2010speech} convolves reverberant speech with the inverse filters to refine speech, but the minimum phase assumption severely restricts this method. Some improvements have been conducted in \cite{wu2006two,mosayyebpour2013single}. But the time-varying RIR is still hard to estimate. (2). Blind multi-channel speech dereverberation is another popular method, which based on multi-channel linear prediction (MCLP) \cite{jukic2017adaptive,jukic2015multi,jukic2016general,yoshioka2009adaptive,yoshioka2013dereverberation}. MCLP-based methods predict undesired reverberant component in the microphone signals, which is subsequently subtracted from the same signals. However, since these methods lack additional knowledge about the undesired components, they may lead to a significant overestimation of these components and severe distortions of the output signals. (3). Spatial processing uses multiple microphones and spectral enhancement to suppress reverberation \cite{allen1977multimicrophone}. Microphone array processing techniques such as beamforming provide spatial filtering to suppress specular reflections so that the speech signals from the desired directions can be enhanced (delay and sum beamformer in \cite{habets2010new}, minimum variance distortionless response (MVDR) beamformer in \cite{habets2013two}). (4). Recently, WPE algorithm has shown promising results \cite{nakatani2008blind,nakatani2010speech}. Standard WPE requires entire utterance to be obtained before calculating the filter taps, and then dereverberation can be performed. Several improvements have also been gradually proposed, such as adaptive version \cite{caroselli2017adaptive} and online WPE \cite{kinoshita2017neural}. WPE and MVDR are also combined in a two stages algorithm \cite{cohen2017combined}. Besides, WPE can be applied to both single and multi-channel conditions.

Nowadays, deep neural network is gradually applied to speech dereverberation. DNN in \cite{han2015learning} learns a spectral mapping from reverberant to anechoic speech. However, performance is still limited at low reverberation time (RT60). Moreover, their system is environmentally insensitive, though LSTM has been utilized to further promote the performance in \cite{weninger2014feature}. Furthermore, DNN-based nonlinear feature mapping and statistical linear feature adaptation approaches are also investigated in \cite{xiao2016speech} for reducing reverberation in speech signals, and a reverberation-time-aware DNN-based speech dereverberation framework \cite{wu2017reverberation} is proposed to handle a wide range of reverberation times. However, this network highly relies on an accurate estimation of RT60.

However, the methods mentioned above come with several shortcomings: (1). Most of the methods rely on complex pipelines composed of multiple algorithms and hand-designed processing stages. (2). MSE loss utilized in these methods only concerns the numerical difference in the estimation, and the numerical error reduction may not necessarily lead to perceptual improvement on the dereverberated speech. (3). Most of methods above only work well in specific environments. Thus we propose a speech dereverberation system based on generative adversarial network. In this paper, we mainly focus on single-channel speech dereverberation. Our contributions are as follows: (1). A more sophisticated structure, CBLDNN, is used to improve the performance; (2). Our system has the ability to deal with the speech with a wide range of RT60 in different environments. (3). GAT is employed to train the network to further enhance the speech quality.

In order to illuminate the effectiveness of this method, several experiments are conducted. Perceptual evaluation of speech quality (PESQ) \cite{rix2001perceptual} and speech
to reverberation modulation energy ratio (SRMR) \cite{falk2008non} are utilized to evaluate the performance. Compared with WPE and deep neural network-based methods, experiments show that our DeReGAT achieves better performances both in PESQ and SRMR. At the same time, our DeReGAT also works well in clean condition, which indicates that our system is robust in variate RT60 and room sizes. Furthermore, we extend the DeReGAT to online speech dereverberation, which enables the output of the proposed method can be thrown into subsequent online speech enhancement and recognition.

The rest of this paper is organized as follows. Section \ref{s2} introduces the single-channel speech dereverberation. Section \ref{s3} describes the generative adversarial training methodology in this paper. Experimental setup and results are presented in Section \ref{s4}. Section \ref{s5} details the online version of the proposed DeReGAT. Finally, Section \ref{s6} concludes our work.

\section{Single-channel speech dereverberation}
\label{s2}
Speech dereverberation aims at estimating anechoic speech signals from reverberant speech. In this paper, we focus on single-channel speech dereverberation task, and additive noise is ignored. The reverberant speech can be represented as:
\begin{equation}
  y(n)=x(n)*h,
  \label{reverb}
\end{equation}
where $y(n)$ represents reverberant speech, $x(n)$ is anechoic speech and $h$ is RIR. The following relationship is still satisfied after short time fourier transform (STFT):
\begin{equation}
  Y(t,f)=X(t,f)\times H,
  \label{frequencySum}
\end{equation}
where $Y(t,f)$ and $X(t,f)$ represent the STFT of speech $y(n)$ and $x(n)$ respectively. Thus, our task is clarified as recovering anechoic signal $x(n)$ from $y(n)$ or $Y(t,f)$. In speech separation task, better results can be obtained by estimating a set of masks \cite{kolbaek2017multitalker}. Similarly, mask is adopted as training target rather than spectral magnitude in our task. In our experiment, we firstly deploy a deep neural network to estimate mask $M(t,f)$ in frequency domain instead of directly recovering $x(n)$ from $y(n)$. In the following equation, $H(|Y(t,f)|,\theta)$ represents a non-linear representation from STFT spectral magnitude $|Y(t,f)|$ to $M(t,f)$.
\begin{equation}
  H(|Y(t,f)|,\theta)=M(t,f),
  \label{Htrans}
\end{equation}
and $|X(t,f)|$ can be recovered by $M(t,f) \times |Y(t,f)|$  ($\times$ indicates element-wise multiplication). The anechoic speech signal $x(n)$ can be obtained by inverse STFT.

Masks are to be estimated as the training targets, and several widely-accepted masks are utilized in paper \cite{kolbaek2017multitalker}. Phase sensitive mask (PSM) takes the differences of phases into account, and it achieves the state-of-the-art performance in speech separation task. Therefore, PSM mask is adopted in this paper, which is defined as:
\begin{equation}
  M^{\mathit{PSM}}=|X(t,f)|\times cos(\theta_{y}(t,f)-\theta(t,f))/|Y(t,f)|,
  \label{PSM}
\end{equation}
where $\theta_{y}(t,f)$ is the phase of reverberant speech, and $\theta(t,f)$ is the phase of clean signal.

\section{Generative adversarial training}
\label{s3}

Generative adversarial net \cite{goodfellow2014generative} comprises two adversarial sub networks, a generator which generates the fake examples from the random noises, and a discriminator which discriminates whether the input is real or generated by the generator.

\begin{figure}[t]
  \centering
  \includegraphics[width=\linewidth]{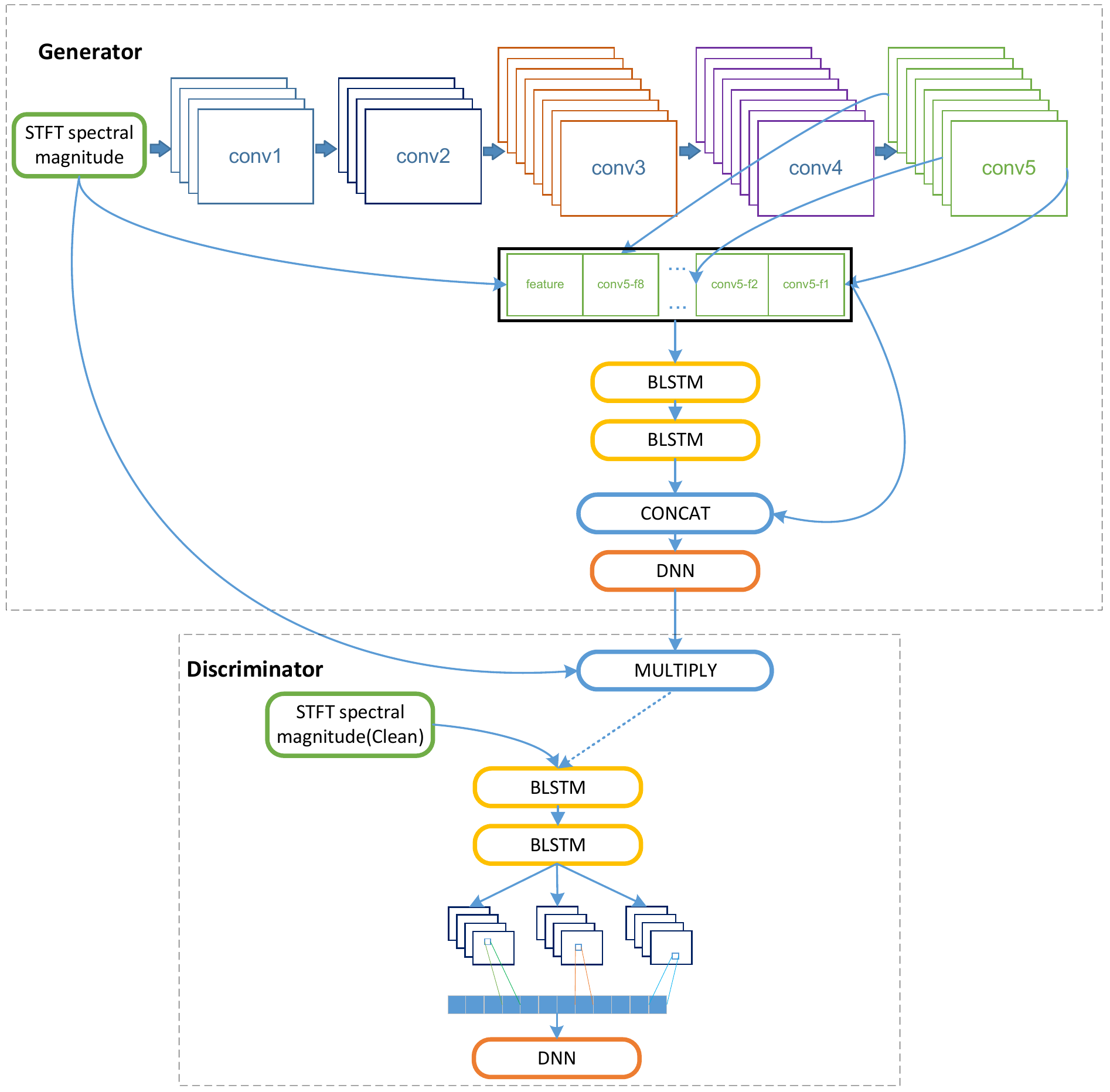}
  \caption{CLDNN-based speech dereverberation with GAT.}
  \label{fig:cldnngan}
\end{figure}

In this paper, we implement a conditional GAN \cite{isola2016image,pascual2017segan}, where the generator, a CLDNN-based model, performs mapping conditioned on some extra clues. Specifically, the generator learns a mapping from observed feature $|Y(t,f)|$ to mask $M(t,f)$. The discriminator is trained to classify whether the STFT feature comes from clean speech or generated. This training procedure is diagrammed in Figure~\ref{fig:cldnngan}.

\subsection{CBLDNN-based generator}

In this section, an utterance-level CLDNN-based generator is proposed. By using convolutional layers, some acoustic variations can be effectively normalized and the resultant feature representation may be immune to speaker variations, colored background and channel noises. Besides, filters that work on local frequency region provide an efficient way to represent local structures and their combinations, which give a more precise spectral structure to dereverberated speech. Higher convolutional layers capture wider frequency variations and cover longer range RT60. BLSTMs well model temporal variations and DNN layers map features to a more separable space. The CBLDNN architecture incorporates the three layers in a unified framework, fusing the benefits of individual layers.

The proposed CBLDNN model is similar to \cite{sainath2015convolutional}, but with fine adjustment and more sophisticated structure. As depicted in Figure~\ref{fig:cldnngan}. The proposed model consists of 5 stacked convolutional layers, 2 stacked BLSTM layers and 1 output layer. In (inChannel, outChannel, kernelW, kernelH) format, the convolutional part have (1, 4, 10, 10)-, (4, 4, 5, 5)-, (4, 8, 7, 7)-, (8, 8, 5, 5)- and (8, 8, 3, 3)-convolution layer from shallow to deep with no pooling and 1 stride. Each BLSTM layer has 256 units. The model has 1 fully-connected (FC) layers, which has 257 output nodes.

\subsection{BLCDNN-based discriminator}
In our experiment, we use utterance-level BLCDNN-based discriminator, which is depicted in Figure~\ref{fig:cldnngan}.

We aim to use BLSTM to model the dependency of the speech and convolutional layer to extract discriminative features that are useful for classification task. As for convolutional layer, to extract complementary features and enrich the representation, we learn several different filters simultaneously. Convolutional filters with multiple sizes capture valuable features from different scales, which contribute a lot to robust classification. The feature maps produced by the convolution layer are forwarded to the pooling layer. 1-max pooling is employed on each feature map to reduce it to a single but the most dominant feature. The features are then joined to form a feature vector input to the final layer. This step transforms the variable-length, high-dimensional vector into a fixed-length vector. Finally, an FC layer maps it to one output node. The input is more like a clean speech when the output value is closer to 1.

The BLCDNN model consists of 2 BLSTM layers, 1 convolutional layer and 1 FC layer. Each BLSTM layer has 256 units. In (kernelW, kernelH) format, the convolutional layer has 3 different filter sizes that are (5, 5), (3, 3) and (1, 1) both with 4 output channels and 1 stride.

\subsection{Loss function}
\label{ssec:loss}

For comparison, a dereverberation system with PSM-based deep neural network (CBLDNN) is selected as baseline, and MSE-based loss function is:
\begin{equation}
  \mathcal{L}_{2}^{\mathit{PSM}} =\frac{1}{N}||M^{\mathit{PSM}}\times |Y|-|X|\cos(\theta_{y}-\theta)||^{2},
  \label{PSM loss}
\end{equation}
where $M$, $|Y|$ and $|X|$ represent mask, STFT spectral magnitude of reverberant speech and STFT spectral magnitude of clean speech for one utterance respectively. $N$ is the total number of time-frequency bins. $\theta_{y}$ is the phase of reverberant speech, and $\theta$ is the phase of clean signal.

In this experiment, we train network by GAT, and LSGAN \cite{mao2016least} based method is utilized. At the same time, $L_1$-regularization is utilized to guide the training. In order to balance GAN loss and $L_1$-regularization, $\lambda$ is taken as hyper-parameter in this experiment. The loss function is listed as:

\begin{equation}
\begin{aligned}
\min\limits_{D} \mathcal{L}(D)&=\mathbb{E}_{|X|\sim{} p_{\mathit{data}}(|X|)}[(D(|X|)-1)^2]\\
                                &+\mathbb{E}_{|Y|\sim{} p_{\mathit{data}}(|Y|)}[(D(G(|Y|)\times |Y|))^2], \\
\min\limits_{G}\mathcal{L}(G)&=\mathbb{E}_{|Y|\sim{} p_{\mathit{data}}(|Y|)}[(D(G(|Y|)\times |Y|)-1)^2]\\
                                &+\lambda\mathcal{L}_{1}^{\mathit{PSM}}.
\label{ganloss}
\end{aligned}
\end{equation}

\section{Experiments}
\label{s4}

\subsection{Experimental setup}
\label{ssec:setup}

To build a large-scale training dataset for DeReGAT, clean speech is selected from WSJ0 dataset \cite{garofalo2007csr}, which has 12776 utterances, 1206 utterances and 651 utterances in training set, development set and test set respectively. Then reverberant speech is generated by convolving the clean speech with RIR. The impulse responses are generated by \cite{allen1979image,lehmann2008prediction}.

We place 1 microphone at the very centre of the room. Three different Room (Room A, B, C) and two different Room (Room D, E) are conducted for training and test. For training and development set, RT60 is uniformly distributed between 0 and 700ms, and RT60 is uniformly distributed between 70ms and 600ms in test set. Sound source is randomly placed in each room. Thus, 38328, 3618 and 1302 utterances are conducted as training set, development set and test set respectively. Detailed configuration is listed in Table~\ref{tab:datatr} and Table~\ref{tab:datatt}. Development set is used to choose the model. The experimental results are all evaluated on test set.

\begin{table}[t]
\caption{\label{tab:datatr} {\it Configurations used for simulating training and development data.}}
\centering
\begin{tabular}{c|c}
\hlineB{3}
\multirow{2}{*}{Dataset} & 12776 utterances in training set and \\
                         & 1206 in development set \\
\hline
\multirow{2}{*}{Room Size(m)} & Room A ($3\times3\times3$), Room B \\
                              & ($6\times6\times4$), Room C ($9\times9\times5$) \\
\hline
$RT_{60}$(ms)  &Uniformly sampled from 0 to 700ms \\
\hlineB{3}
\end{tabular}
\end{table}

\begin{table}[t]
\caption{\label{tab:datatt} {\it Configurations used for simulating test data.}}
\centering
\begin{tabular}{c|c}
\hlineB{3}
Dataset & 651 utterances in test set \\
\hline
\multirow{2}{*}{Room Size(m)} & Room D ($4\times5\times3$), Room E \\
                              & ($10\times12\times6$) \\
\hline
$RT_{60}$(ms)  & Uniformly sampled from 70ms to 600ms \\
\hlineB{3}
\end{tabular}
\end{table}

The input features of generator and discriminator are 257-dimensional STFT spectral magnitude computed with a frame size of 32ms and 16ms shift. The phase of the anechoic signal is used to build PSM-based loss function, and the phase of the reverberant speech is used to restore the speech. After fine adjustment, hyper-parameters $\lambda$ is set as 1. The models are all trained on Tensorflow \cite{abadi2016tensorflow}. RMSprop algorithm \cite{tieleman2012lecture} is utilized for training where the learning rate started at 0.0002.

\subsection{Baseline systems}
\label{ssec:resultbaseline}

We conduct several baseline systems. A standard WPE-based system and a deep neural network-based system, CBLDNN, are conducted. CBLDNN outputs PSM-based mask, and MSE is selected as loss function. CBLDNN has the same model structure as the generator of DeReGAT. The performance of these systems is listed in Table~\ref{tab:GAT}.

\subsection{DeReGAT dereverberation system}
\label{ssec:resultMTL}

\begin{figure*}[!htb]
\centering
\subfloat[Reverberant speech.]{\includegraphics[width=.24\textwidth]{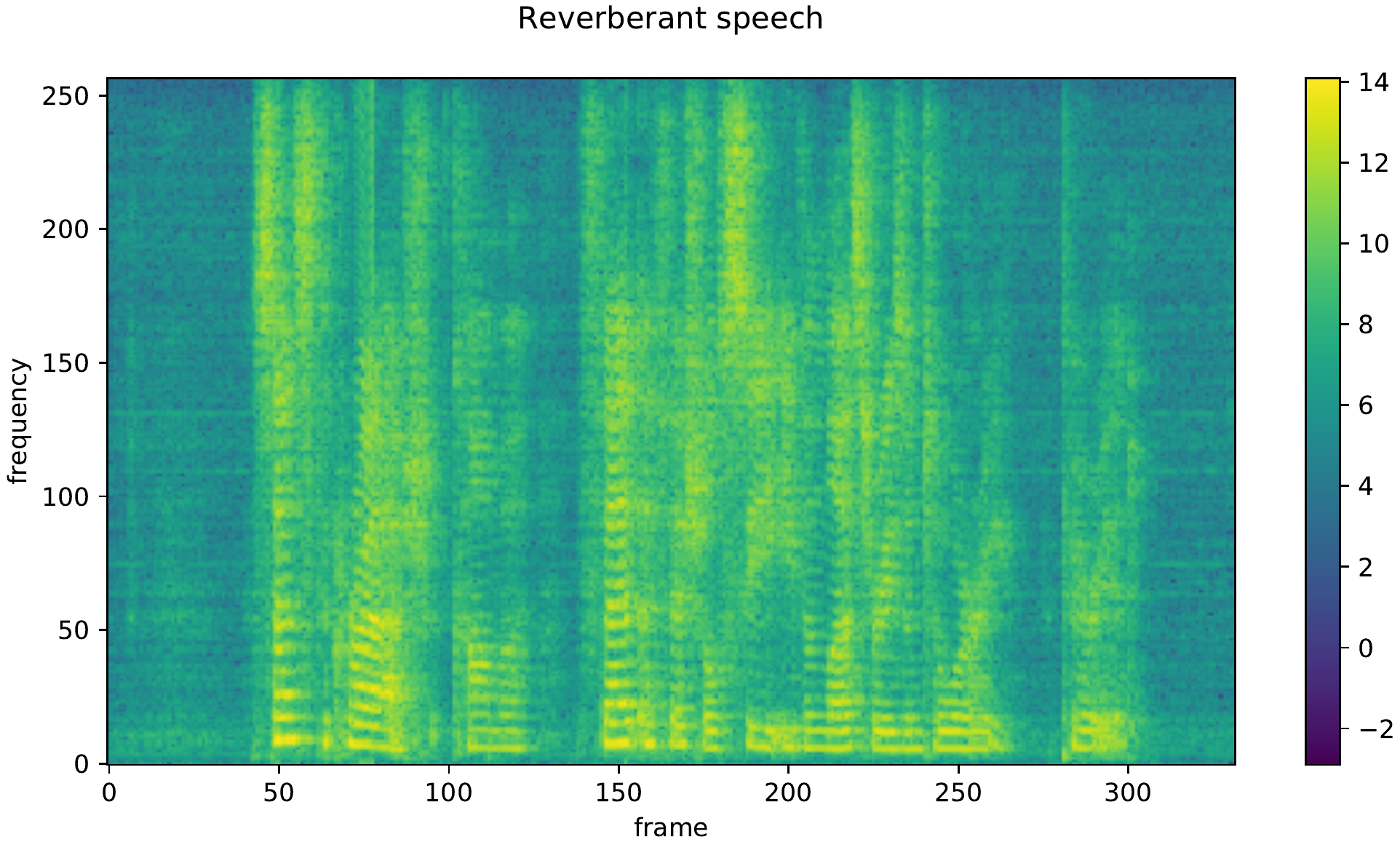}}\hfill
\subfloat[Dereverberated by WPE.]{\includegraphics[width=.24\textwidth]{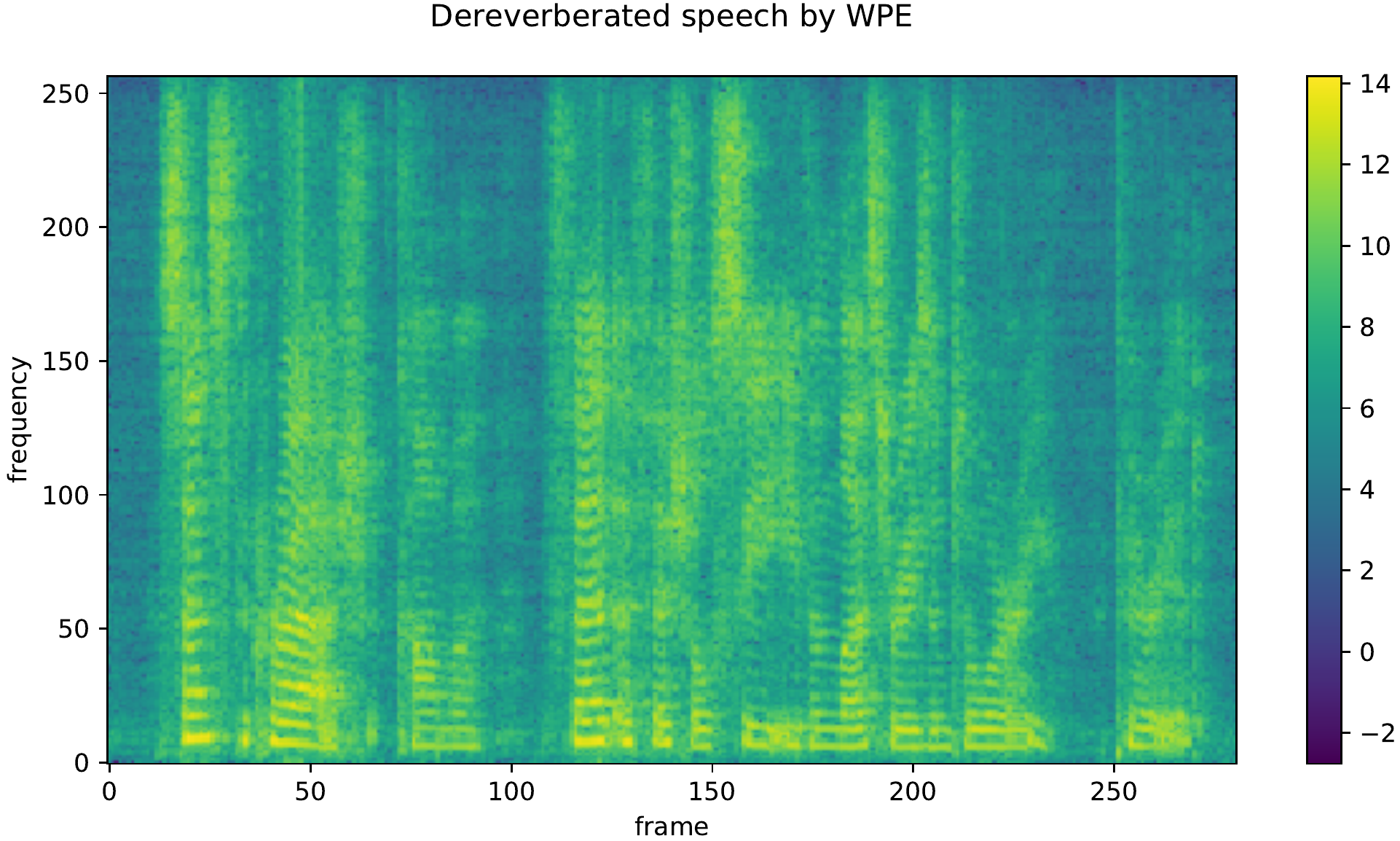}}\hfill
\subfloat[Dereverberated by CLDNN.]{\includegraphics[width=.24\textwidth]{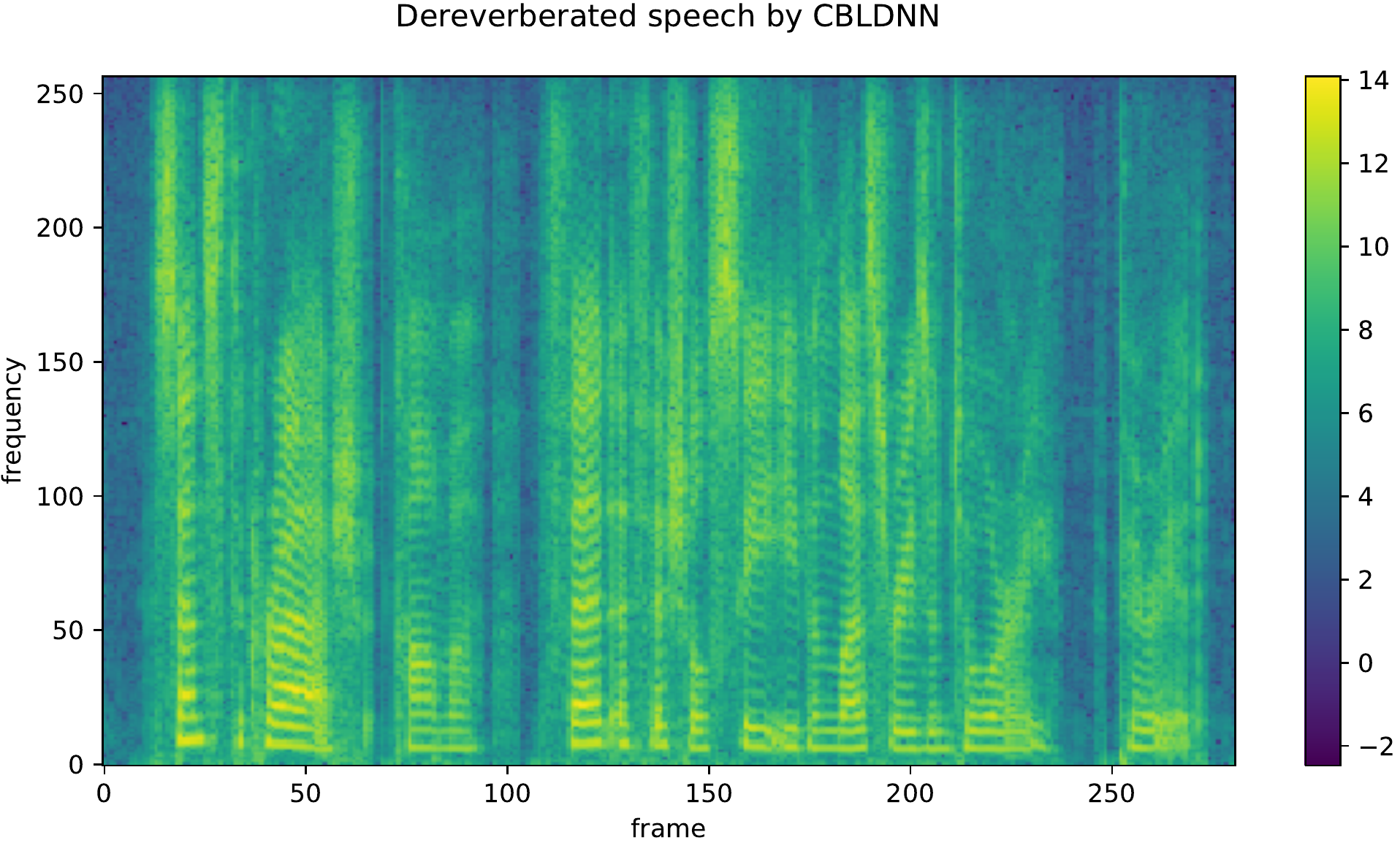}}\hfill
\subfloat[Dereverberated by DeReGAT.]{\includegraphics[width=.24\textwidth]{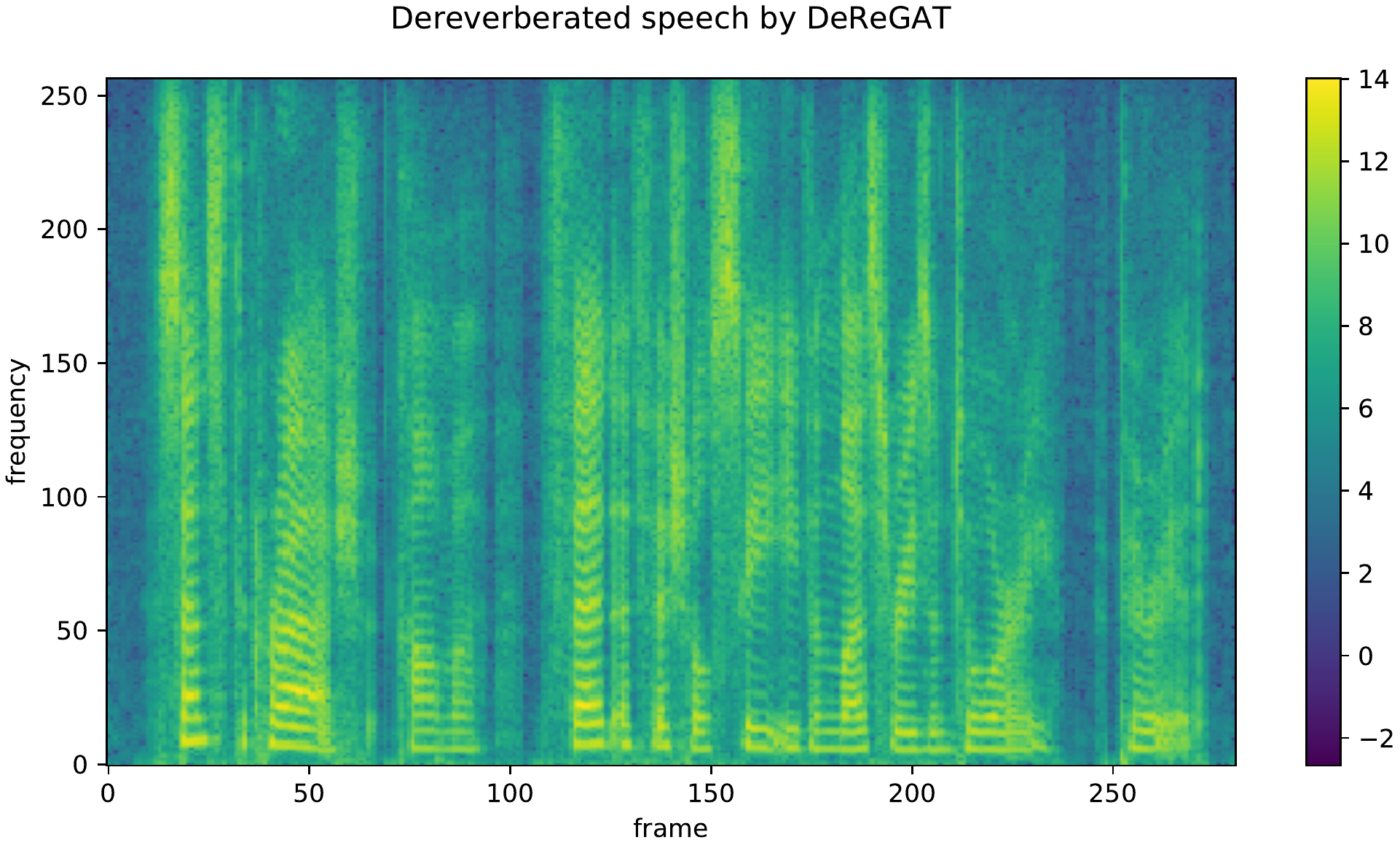}}\hfill
\subfloat[Dereverberated by DeRe-10.]{\includegraphics[width=.24\textwidth]{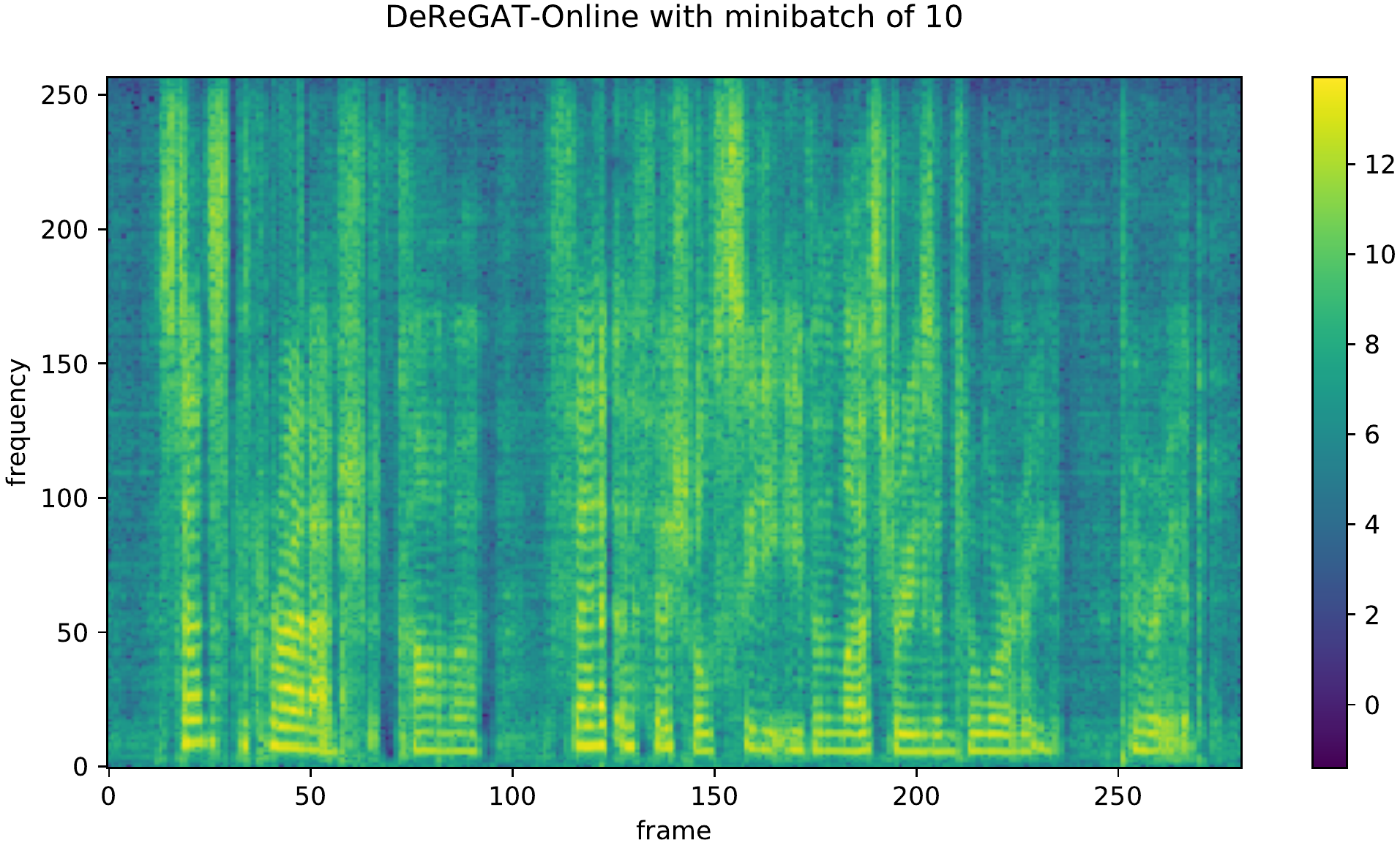}}\hfill
\subfloat[Dereverberated by DeRe-20.]{\includegraphics[width=.24\textwidth]{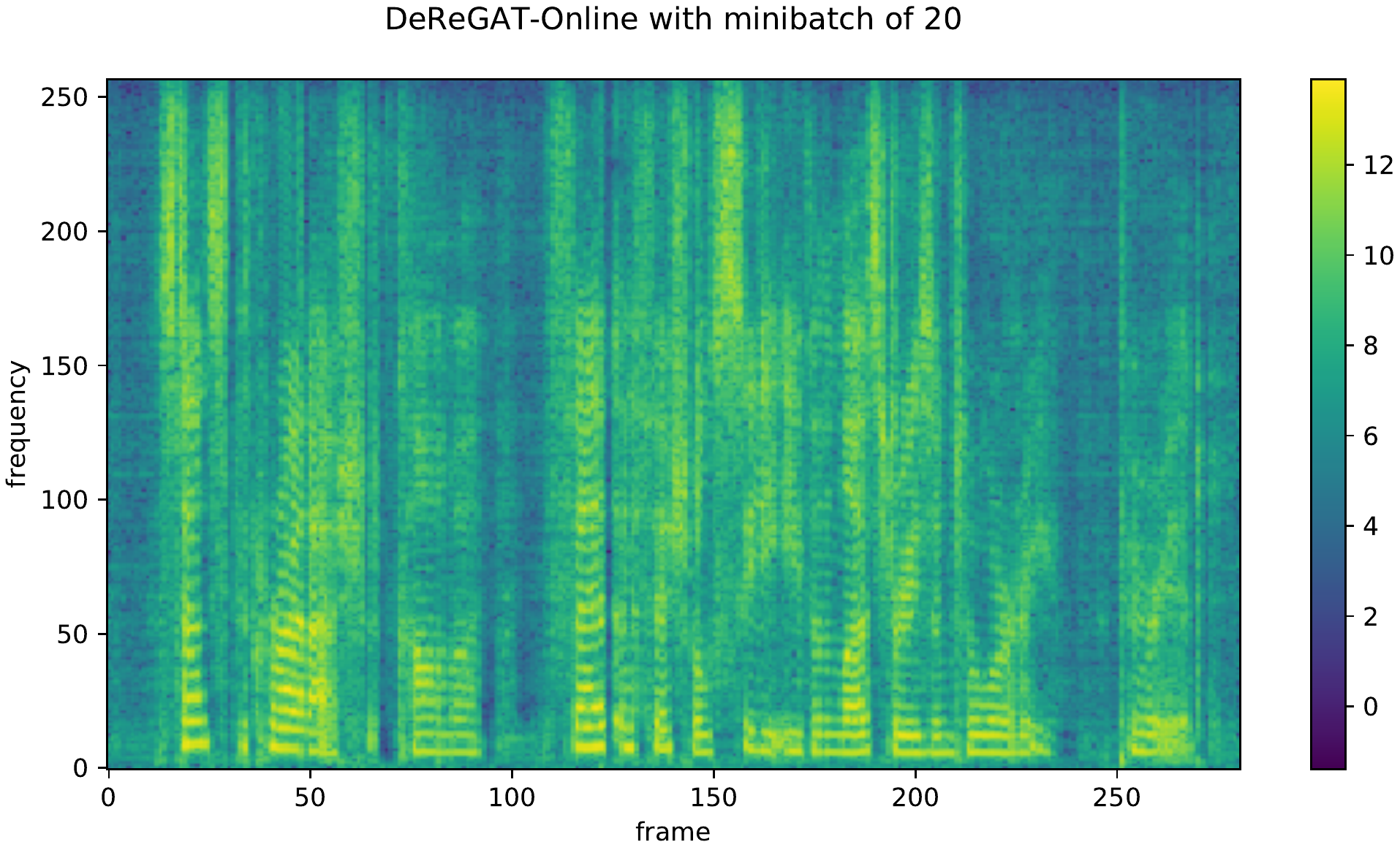}}\hfill
\subfloat[Dereverberated by DeRe-40.]{\includegraphics[width=.24\textwidth]{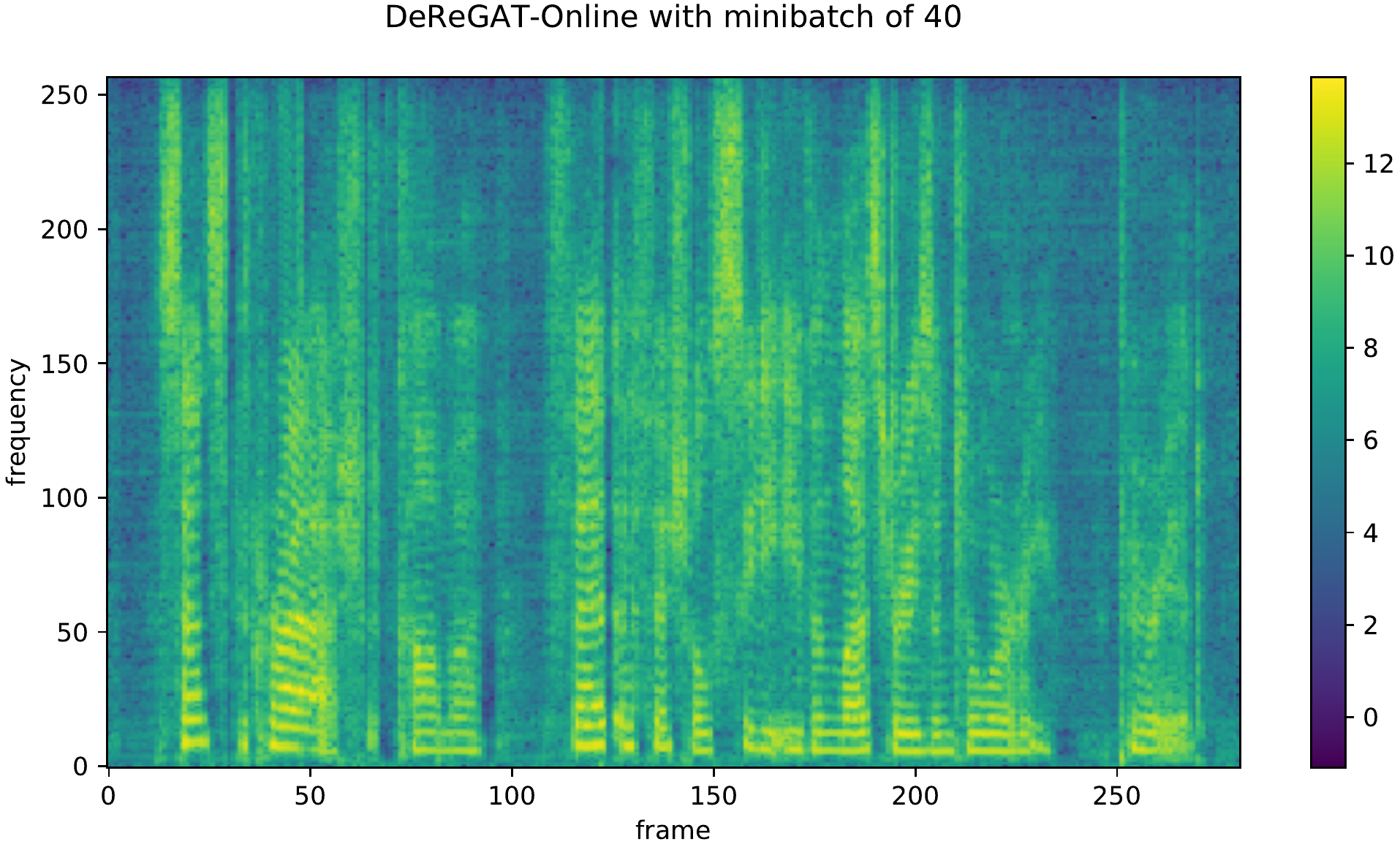}}\hfill
\subfloat[Clean signal.]{\includegraphics[width=.24\textwidth]{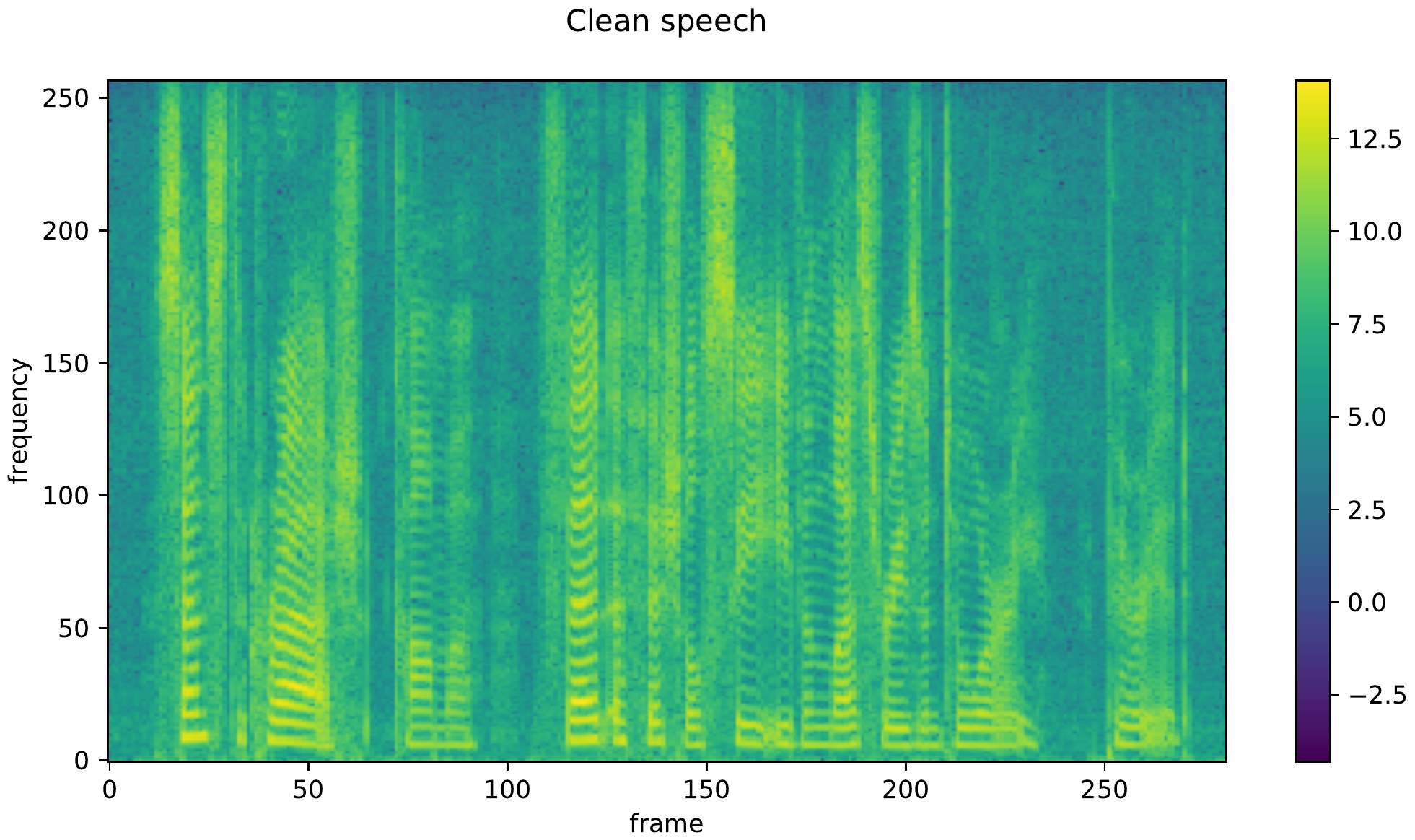}}
\caption{An example of dereverberated speech and its clean signal, which is randomly selected. The RT60 of reverberant speech is 468ms.}
\label{f2}
\end{figure*}

\begin{table}[t]
\newcommand{\tabincell}[2]{\begin{tabular}{@{}#1@{}}#2\end{tabular}}
\caption{\label{tab:GAT} {\it Comparison of dereverberation systems with different methods.}}
\centering
\renewcommand{\multirowsetup}{\centering}
\scalebox{0.86}{
\begin{tabular}{c c c c c c c}
\hlineB{3}
\multirow{2}*{System} & ~ &\multicolumn{2}{c}{PESQ} & ~ &\multicolumn{2}{c}{SRMR}  \\
\cline{3-4}
\cline{6-7}
& ~ & Room D & Room E & ~ & Room D & Room E \\
\hline
\hline
WPE & ~ & $2.00$ & $2.21$ & ~ & $4.45$ & $4.67$ \\
CBLDNN & ~ & $2.28$ & $2.37$ & ~ & $4.93$ & $5.11$ \\
DeReGAT & ~ & $2.54$ & $2.76$ & ~ & $5.70$ & $5.42$ \\
\hline
\hline
Reverberant & ~ & $1.75$ & $2.04$ & ~ & $4.09$ & $4.28$ \\
Clean & ~ & $-$ & $-$ & ~ & $5.73$ & $5.62$ \\
\hlineB{3}
\end{tabular}
}
\end{table}
In this section, we explore the impact of GAT. In GAT, PSM is used to measure the loss of $L_1$, which is utilized to minimize the distance between generations and clean examples. From Table~\ref{tab:GAT}, experimental results show that all listed methods have contribution to dereverberation in different rooms, and DeReGAT obtains the best results. Compared with existing methods, PESQ and SRMR of DeReGAT have been significantly improved. The SRMR of speech processed by DeReGAT is markedly approximate to the clean speech. The results mean GAT plays an important role in improving the performance. GAT makes the speech produced by generator approaches to clean one. The results also state that our DeReGAT achieves better dereverberation performance in variant environments.

Figure~\ref{f2} shows the spectrogram of dereverberated speech based on different methods. Compared with speech dereverberated by WPE and CBLDNN, the speech dereverberated by DeReGAT has clearer and more compact spectrum structure. At the same time, the speech dereverberated by WPE remains obviously reverberations.

\subsection{DeReGAT in clean environment}
\label{ssec:resultGAT}

In practice, due to environments differences, speech collected by microphone may bear various degrees of reverberation or have little reverberation. Therefore, We augment clean samples and weakly-reverberated samples into training set to ensure that dereverberation system not only has an impressive effect to dereverberate reverberant speech but also maintain speech with zero or little reverberation. In the testing phase, clean speech which derived from different environments is sent to DeReGAT. The results are shown in Table~\ref{tab:clean}. From the view of PESQ, clean speech is distorted by WPE apparently, but it still shows high quality after passing DeReGAT both in Room D and Room E. This shows that our system can be applied to different environments with variant reverberation.

\begin{table}[t]
      \newcommand{\tabincell}[2]{\begin{tabular}{@{}#1@{}}#2\end{tabular}}
        \caption{\label{tab:clean} {\it Comparison of dereverberation systems with clean input.}}
        \vspace{2mm}
        \centerline{
          \renewcommand{\multirowsetup}{\centering}
          \begin{tabular}{c c c}
            \toprule[1pt]
            \multirow{2}*{System}  &\multicolumn{2}{c}{PESQ}  \\
            \cline{2-3}
            & Room D & Room E\\
            \hline
            \hline
              WPE  & $3.44$      & $3.11$   \\
              DeReGAT         & $4.47$      & $4.48$        \\
            \bottomrule[1pt]
          \end{tabular}
        }
\end{table}

\section{Online speech dereverberation by DeReGAT}
\label{s5}
\begin{table}[ht]
\newcommand{\tabincell}[2]{\begin{tabular}{@{}#1@{}}#2\end{tabular}}
\caption{\label{tab:online} {\it Comparison of dereverberation systems with online processing.}}
\centering
\renewcommand{\multirowsetup}{\centering}
\scalebox{0.86}{
\begin{tabular}{c c c c c c c}
\hlineB{3}
\multirow{2}*{System} & ~ &\multicolumn{2}{c}{PESQ} & ~ &\multicolumn{2}{c}{SRMR}  \\
\cline{3-4}
\cline{6-7}
& ~ & Room D & Room E & ~ & Room D & Room E \\
\hline
\hline
DeRe-10 & ~ & $1.54$ & $1.55$ & ~ & $4.96$ & $4.77$ \\
DeRe-20 & ~ & $2.12$ & $2.23$ & ~ & $5.44$ & $5.25$ \\
DeRe-40 & ~ & $2.33$ & $2.41$ & ~ & $5.63$ & $5.35$ \\
\hline
\hline
DeReGAT & ~ & $2.54$ & $2.76$ & ~ & $5.70$ & $5.42$ \\
Reverberant & ~ & $1.75$ & $2.04$ & ~ & $4.09$ & $4.28$ \\
\hlineB{3}
\end{tabular}}
\end{table}

In this section, we extend the proposed DeReGAT to enable online speech dereverberation (DeReGAT-online). We assume that the reverberant signal is obtained as a sequence of mini-batches. In detail, compared with offline DeReGAT, DeReGAT-online is performed without updating parameters. DeReGAT-online receives the mini-batch and outputs the dereverberated batch. The size of the mini-batches are selected as 10 (DeRe-10), 20 (DeRe-20), 40 (DeRe-40) frames, which represent 160ms, 320ms, 640ms delays respectively.

The results are listed in Table~\ref{tab:online}. Compared with utterance-level DeReGAT, the performance of DeReGAT-online degrades. Compared with reverberant speech, the worse PESQ of DeRe-10 shows heavy distortion in speech quality. The comparable performance of DeRe-40 with DeReGAT makes it capable of serving as an online system, where the overall delay consists of model's feed-forward time and the size of mini-batch. An example is also depicted in Figure~\ref{f2}. The visual intervals in Figure~\ref{f2}(e) make the corresponding speech sounds off and on. The speech dereverberated by DeRe-20 sounds better, and DeRe-40 generates the most consistent speech among all online systems.

\section{Conclusions}
\label{s6}
In this paper, we introduce CBLDNN-based single-channel speech dereverberation system with GAT. Our results indicate that DeReGAT shows impressive performance improvement compared with traditional WPE and MSE-based deep neural network methods. Our DeReGAT effectively deals with variant RT60 and environmental differences. Additionally, DeReGAT is extended to an online version, which can achieve comparable performance with offline DeReGAT. This shows great potential for further deployment. We note that the proposed method has great potential for the further improvement. We can compress the model to further reduce the system size, which can reduce the storage size and feed-forward time. Furthermore, single-channel DeReGAT can be extended to multi-channel DeReGAT, which can utilize the information of source location.

\section{Acknowledgments}
This work is supported by National Key Research and Development Program of China under Grant No.2016YFB1001404 and Beijing Engineering Research Center Program under Grant No.Z171100002217015.

\bibliographystyle{IEEEtran}
\bibliography{bibtex}

\end{document}